\documentclass[a4paper]{article}
\usepackage{INTERSPEECH2019}

\title{Telephonetic: Making Neural Language Models Robust to ASR and Semantic~Noise}
\name{Chris Larson \quad Tarek Lahlou \quad Diana Mingels \quad Zachary Kulis \quad Erik Mueller}
\address{Capital One}
\email{\{christopher.larson2, tarek.lahlou, diana.mingels, zachary.kulis, erik.mueller \}@capitalone.com}

\def \w{ \mathbf{w} }

\begin{document}

\maketitle

\begin{abstract}
    Speech processing systems rely on robust feature extraction to handle phonetic and semantic variations found in natural language. While techniques exist for desensitizing features to common noise patterns produced by Speech-to-Text (STT) and Text-to-Speech (TTS) systems, the question remains how to best leverage state-of-the-art language models (which capture rich semantic features, but are trained on only written text) on inputs with ASR errors. In this paper, we present {\tt Telephonetic}, a data augmentation framework that helps robustify language model features to ASR corrupted inputs. To capture phonetic alterations, we employ a character-level language model trained using probabilistic masking. Phonetic augmentations are generated in two stages: a TTS encoder (Tacotron 2, WaveGlow) and a STT decoder (DeepSpeech). Similarly, semantic perturbations are produced by sampling from nearby words in an embedding space, which is computed using the BERT language model. Words are selected for augmentation according to a hierarchical grammar sampling strategy. Telephonetic is evaluated on the Penn Treebank (PTB) corpus, and demonstrates its effectiveness as a bootstrapping technique for transferring neural language models to the speech domain. Notably, our language model achieves a test perplexity of 37.49 on PTB, which to our knowledge is state-of-the-art among models trained only on PTB. \\

\end{abstract}

\section{Introduction}
    Language modeling is the core component of both written and spoken language understanding systems. Recent work on transformer networks \cite{Vaswani-AIA, Devlin18-BPO, Radford19-LMA} trained on massive written text corpora have repeatedly demonstrated step changes in performance on several tasks including text summarization, question answering (Q\&A), intent classification, natural language inference (NLI), as well as in \textit{zero-shot} settings, in which they generalize to new tasks and new data without being pre trained on that data or those tasks. Leveraging these models in spoken language understanding systems, however, is obfuscated by the fact that noise injected by automatic speech recognition (ASR) processing components contains structure that is otherwise not taken advantage of. Previously, researchers have developed approaches to learn real-valued representations of spoken language that are sensitive to acoustic and phonetic similarities \cite{Dmitriy18-TEE}, and more recently semantic similarity such as Speech2Vec \cite{Chung18-SAS}. These approaches have clear drawbacks, however, in that they are stand-alone models, they are trained on one task, and they do not leverage the rich, pre trained feature layers offered by models such as BERT \cite{Devlin18-BPO} and GPT-2 \cite{Radford19-LMA}.
    
    In this paper, we introduce the {\it Telephonetic} augmentation framework wherein data augmentations can be used to fine-tune language models on written text data so that they are better equipped to handle ASR errors. The term telephonetic is inspired by the popular game {\it telephone} wherein a message is passed sequentially and orally from person to person, aggregating phonetic and other errors along the way, until eventually the message has diverged significantly from its origin. This paper lays the foundation for generating similar errors by pairing neural speech synthesis systems with commodity ASR systems and reflecting the resulting errors into a training dataset. In Section~2 we discuss the character level language model evaluated in this paper as well as the companion training strategy. In Section~3 we present the core components of the telephonetic framework and provide experimental results and commentary in Section~4.
    
\section{Character-based language modeling}

\begin{figure}[]
        \begin{centering}
            \includegraphics[width=3.0in]{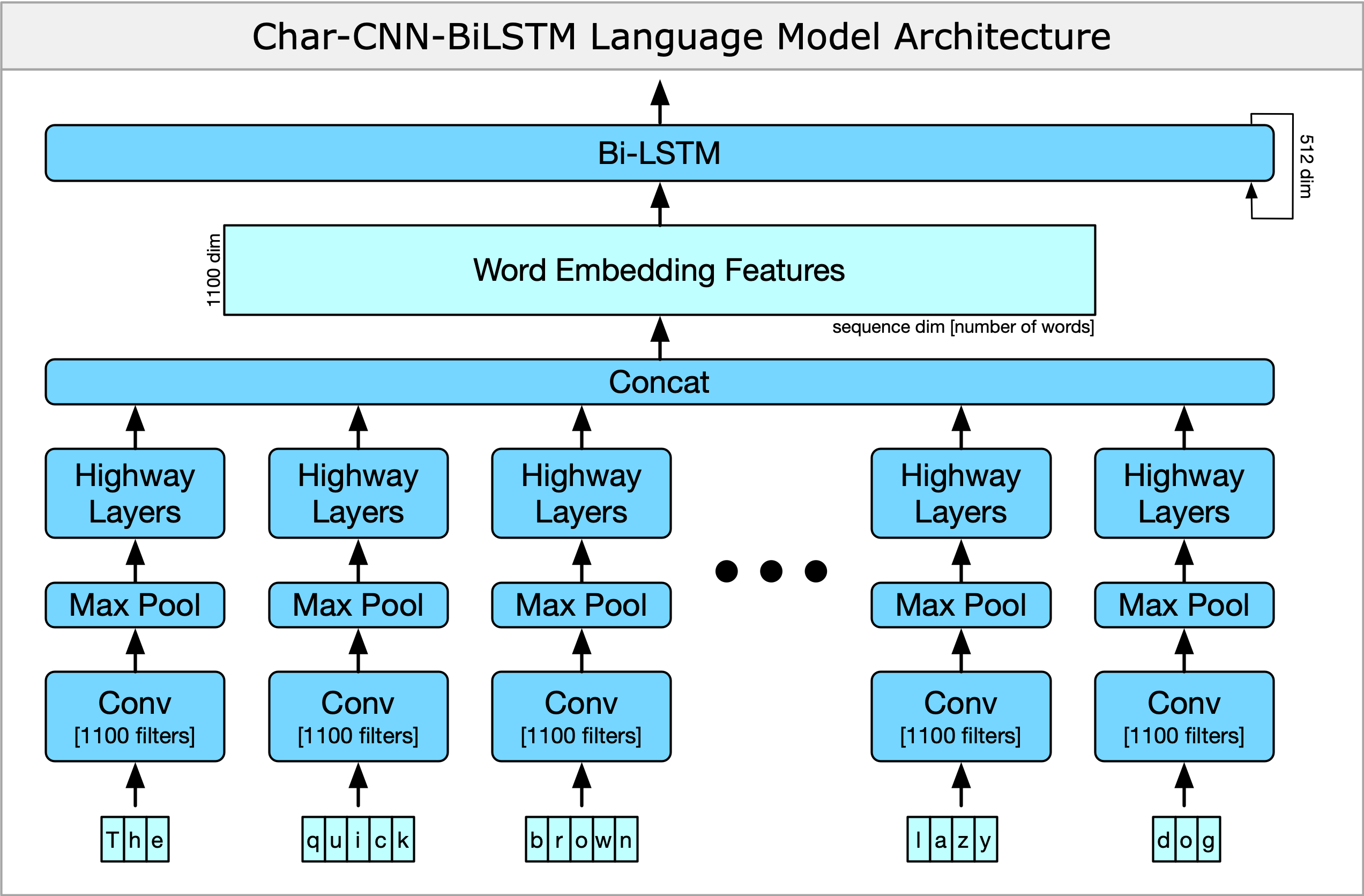}\par
        \end{centering}
        \caption{The Char-CNN-BiLSTM architecture derived from \cite{Kim15-CAN}. Convolutional filters are used to convolve over character embeddings for each word in the input sequence. Two highway transformations are then applied prior to the application of two bi-directional LSTM layers. The model outputs a probability distribution over its word vocabulary.}
        \label{fig:char-cnn}
    \end{figure}

    The Language Model (LM) discussed in this paper builds upon the Char-CNN-LSTM architecture proposed in \cite{Kim15-CAN}. Rather than imposing causality to the learning task, as is commonplace with next-word prediction training, we instead use a masked LM training procedure inspired by Devlin \textit{et al.} \cite{Devlin18-BPO}. Letting $\w_{i}$ denote the $i$-th word in the text sequence $\w = [\w_{0}, \dots, \w_{T-1}]$, of length $T$ the masked LM training strategy is to randomly replace a word $\w_{i}$ with a mask token and then attempt to predict the masked word. The contribution of a single sample to the negative log-likelihood, $-log(\ell)$, training objective is specified according to 
    \begin{eqnarray} \label{eq:mlm-objective}
    	-\log{\ell(\theta_{LM} | \w)} = \displaystyle - \log{p(\w_i | \tilde{\w}_{c}^i; \theta_{LM})},
    \end{eqnarray}
    where $\theta_{LM}$ denotes the LM parameters and $\tilde{\w}_c^i$ denotes the character representation of $\w$ with the $i$-th entry or word having been replaced with the mask token. Importantly, we do not accumulate loss over all token in an example, only one randomly chosen token; this decorrelates the gradients at the mini-batch level. To better align the Char-CNN-LSTM architecture to the masked LM training procedure, we employ a bi-directional LSTM head for the prediction task since masking the label at the input layer allows us to fully recurse both LSTMs over the entire input sequence (including the mask token itself). This modified architecture, which we refer to hereon as Char-CNN-BiLSTM, is depicted in Fig.~\ref{fig:char-cnn}. The training procedure consists of randomly selecting one word from an input text sequence, masking it with probability $p_{m} = 0.85$, and then performing mini-batch stochastic gradient descent using the objective in \eqref{eq:mlm-objective}. We specifically chose to focus attention on language models that operate on character-level tokens rather than the word-piece tokens in \cite{Devlin18-BPO} due to the resilience character level models have to spelling errors in the input text. 
    
    \begin{figure*}[ht]
        \begin{centering}
            \includegraphics[width=6.5in]{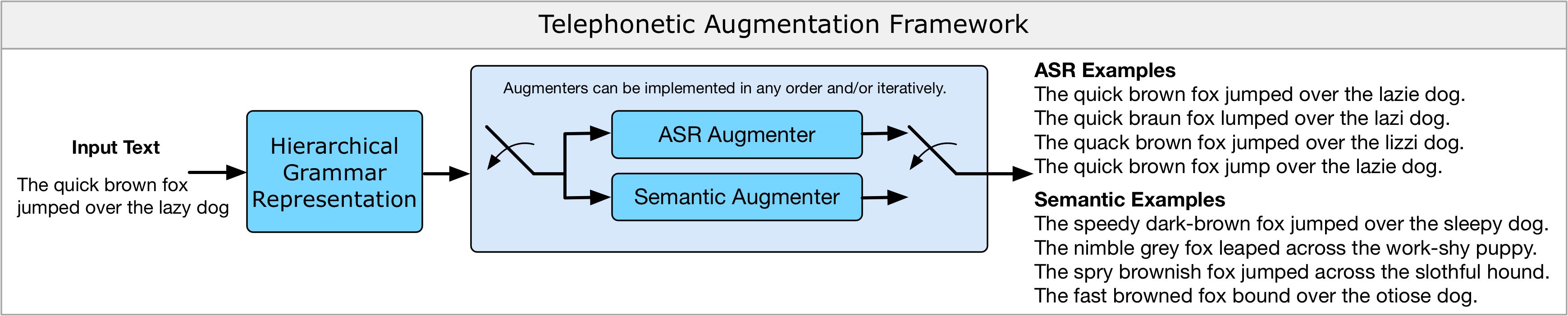}\par
        \end{centering}
        \caption{A high level description of the telephonetic augmentation framework.}
        \label{fig:data-augment}
    \end{figure*}

\section{Augmentation strategy}
    The telephonetic augmentation strategy presented in this work involves sampling text inputs according to their grammar and then augmenting portions of those inputs by running them through noisy ASR and semantic replacement systems. The following steps summarize this procedure for a given input text: 
    \begin{enumerate}
    	\item Express the grammatical hierarchy of the input text using a directed graph. 
        \item Sample and replace nodes in the graph with semantically similar nodes. 
        \item Sample and replace nodes in the graph with the ASR system outputs produced using synthesized audio of the text within the nodes.
    \end{enumerate} 
    Several input-output examples as well as the above steps are illustrated in Fig.~\ref{fig:data-augment}. The individual processing modules that comprise the framework are described in detail in the remainder of this section.  
    
    \subsection{Hierarchical grammar based sampling}
    
        The purpose of the sampling method is to extract portions of the input text to augment that align with the desired robustness metrics and potential use cases of the language model. Toward this goal, we leverage hierarchical representations of the inputs grammar in order to efficiently sample based on part-of-speech by traversing the graph for certain node types. For example, sampling and augmenting different grammatical elements are potentially more suitable to downstream tasks than others, e.g. verbs for intent or sentiment classification. Indeed, sampling along the graph allows for the selection of either individual words or phrases based upon the particular tree description of the text and whether a node or a node's subtree are selected.
        
\begin{figure}[t]
            \begin{centering}
                \includegraphics[width=2.5in]{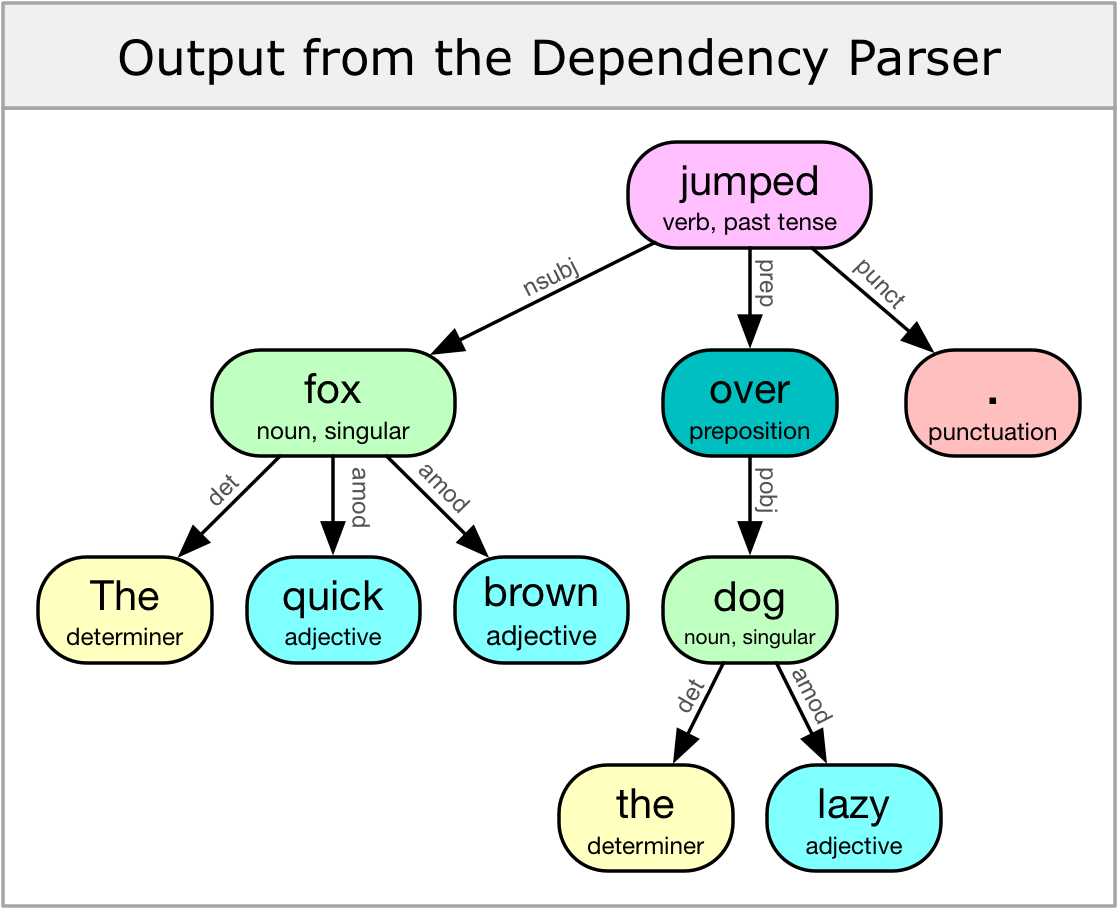}\par
            \end{centering}
            \caption{An example graph representing the grammatical dependencies in the sentence {\tt The quick brown fox jumped over the lazy dog.}}
            \label{fig:dp}
        \end{figure}

        One instance of such a representation can be achieved using a dependency parser. Broadly speaking, dependency parsers expose grammatical relationships between {\it head} words and words which modify said heads thereby providing directed graphs. An example graph is portrayed in Fig.~\ref{fig:dp} for the example input text used in Fig.~\ref{fig:data-augment}. In this paper, we utilize a dependency parser to strategically sample words in the input text which are well-suited to making a language model robust to input texts that draw upon large vocabularies and contain many ASR errors, i.e., we focus on augmentations of adjectives and nouns. 
    
    \subsection{Semantic augmentation}
    
        The goal of semantic augmentation in this context is two-fold. First, it provides a method to sensitize pre-trained language models to domain-specific language through data augmentation; this is useful in the absence of large domain-specific datasets. A second, more subtle benefit applies specifically to the ASR context, in which phonetically corrupted words can be mapped back to semantically similar, in-vocabulary words.
        
        To generate semantically altered text samples, we leverage the BERT language model \cite{Devlin18-BPO}, which was trained on the BookCorpus and English Wikipedia corpora, totalling 3.3B words. BERT employs a multi-embedding input layer that consists of word-piece tokens, positional tokens, and sequence tokens that are summed and then fed into the transfer layers, which consist of dense fully connected layers. The semantic augmentation strategy is illustrated in Fig.~\ref{fig:embed}(a). The core component of the semantic augmenter is a nearest-neighbor look-up table for each word in a lexicon of 80K common English words. Nearest neighbors are found by computing the covariance of the 80K lexicon in the 768-dimensional BERT embedding space, sorting the rows in descending order, and then packing the corresponding sorted words into a look-up table. At train time, the semantic augmenter replaces words simply by sampling uniformly from its top-5 nearest neighbors. 

    \subsection{ASR augmentation}
    
        The primary goal of ASR augmentation is to produce spans of output text that coincide with common transcription errors that ASR systems would make when the input text is voiced by a number of speakers with diverse acoustic, linguistic, and phonetic qualities. The ASR augmenter utilized in this paper is portrayed in Fig.~\ref{fig:asr}(a) and is made up of two parts: a text-to-speech engine and a speech-to-text engine. Referring to the figure, the depicted text-to-speech engine utilizes Tacotron 2 \cite{Wang17-TTE} to produce mel spectrogram representations of an input text that are than fed to WaveGlow \cite{Prenger18-WAF} to produces audio outputs similar to those recorded by humans. In more detail, the Tacotron 2 architecture is depicted in Fig.~\ref{fig:asr}(b) and maps character embeddings through a sequence-to-sequence network with attention to a sequence of mel spectrogram frames. WaveGlow, whose architecture is depicted in Fig.~\ref{fig:asr}(c),  is a state-of-the-art flow-based network able to synthesize high quality audio from mel spectrograms. As a consequence of the single-speaker identity of each architecture, every speaker identity requires a unique pair of trained Tacotron 2 and WaveGlow models.  The speech-to-text engine depicted utilizes DeepSpeech \cite{Kim15-CAN} to transform the generated audio back into text.
        
\section{Experiments}
        
    \subsection{ASR augmenter training}
    
\begin{figure}[t]
        \begin{centering} 
            \includegraphics[width=3in]{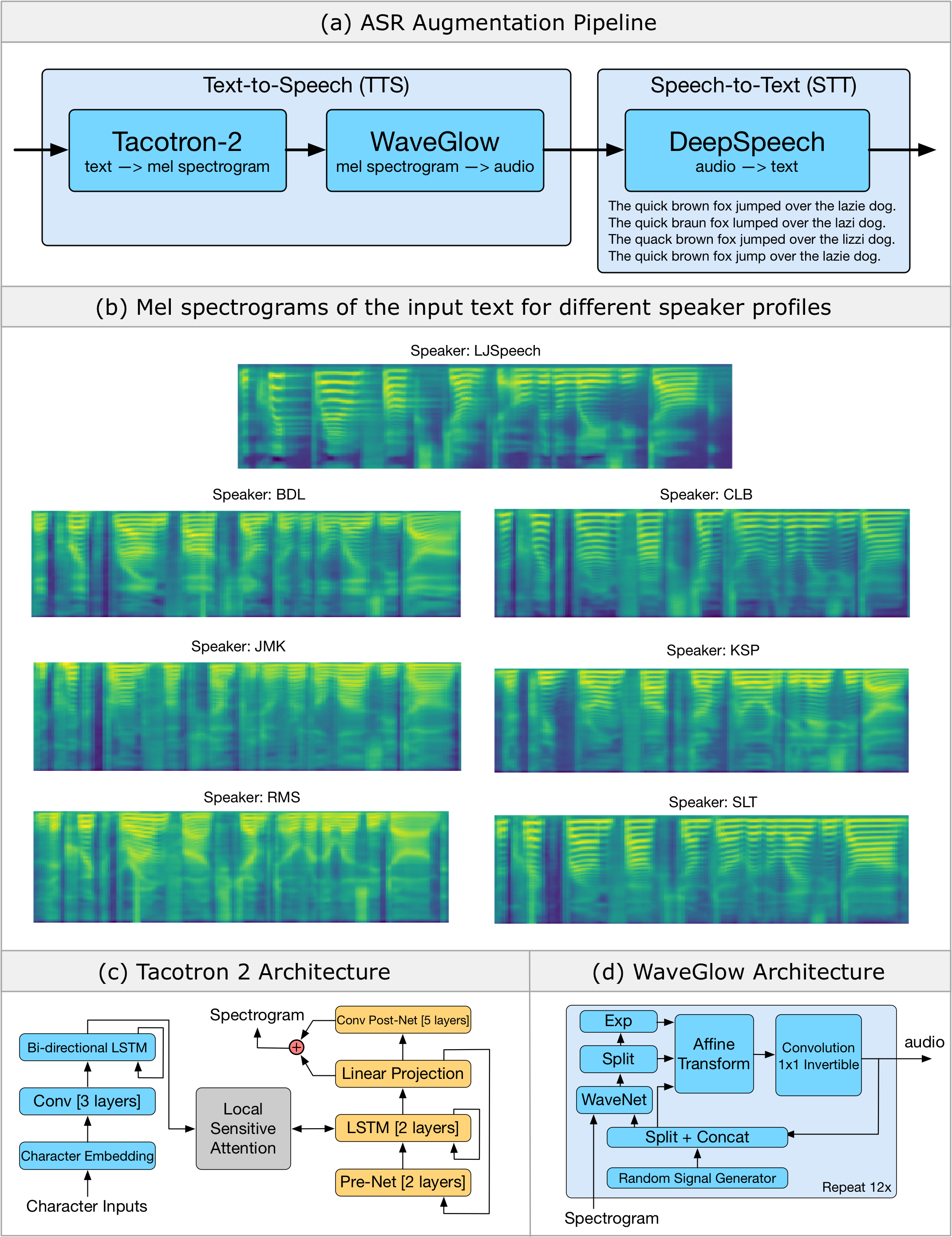}\par
        \end{centering}
        \caption{(a) An example ASR augmentation pipeline with a text-to-speech engine depicted using Tacotron 2 and WaveGlow and a speech-to-text engine depicted using DeepSpeech. (b) The mel spectrogram representations of the example text from Fig.~\ref{fig:data-augment} for the LJSPeech dataset (22kHz) and CMU speakers (16kHz). The architectural details of (c) Tacotron 2 and (d) WaveGlow are also provided.}
        \label{fig:asr}
    \end{figure}

        The ASR augmenter used in the following experiments was built by training the text-to-speech engine depicted in Fig.~\ref{fig:asr}(a) for each of the speaker profiles in the CMU Arctic dataset \cite{Kominek03cmuarctic}. These profiles contain five male and two female voices, four of which have US English accents and one of each Canadian, Indian and Scottish English accents. Since each profile only contains approximately one hour of transcribed and aligned data, we first train both the Tacotron 2 and WaveGlow models on the LJSpeech dataset \cite{ljspeech17} which contains 13,100 short audio clips totaling approximately 24 hours of a single speaker reading passages from 7 non-fiction books. Both of these models were each trained on 8 Tesla V100 GPUs for several days until audio samples containing out of vocabulary words qualitatively sounded natural. Both models used the ADAM optimizer; Tacotron 2 used a batch size of 128, weight decay of 1e-6, and an initial learning rate of 1e-3, whereas WaveGlow used a batch size of 32, no weight decay, and an initial learning rate of 1e-4. 

        For each CMU speaker, a Tacotron 2 and WaveGlow model was then fine-tuned from the LJSpeech models on 4 Tesla V100 GPUs for a number of additional days until comparable quality in the new voice was achieved. Both models fine tuned using the ADAM optimizer with an initial learning rate of 1e-5 and weight decay of 1e-6. Tacotron 2 used a batch size of 12 whereas WaveGlow used a batch size of 8. For the fine-tuned Tacotron 2 models, the sampling rate mismatch between the two datasets required the alignment mechanisms to be relearned from scratch. All models trained for the text-to-speech engines used manual learning rate annealing. 
        
        While the two models for a given speaker profile were not trained end-to-end, the mel spectrograms used as inputs to training WaveGlow were produced by the fully trained, corresponding Tacotron 2 model rather than derived from the original audio files. In this sense, WaveGlow has the opportunity to learn to correct for spectral errors systematically produced by Tacotron 2. Figure~\ref{fig:asr}(b) illustrates the final mel spectrograms for the example sentence in Fig.~\ref{fig:data-augment} for the LJSpeech speaker and six of the CMU Arctic speakers. Note that the sampling rate for the LJSpeech audio is higher which manifests itself through the longer mel spectrum.          
        
    \begin{figure}[t]
        \begin{centering}
            \includegraphics[width=3in]{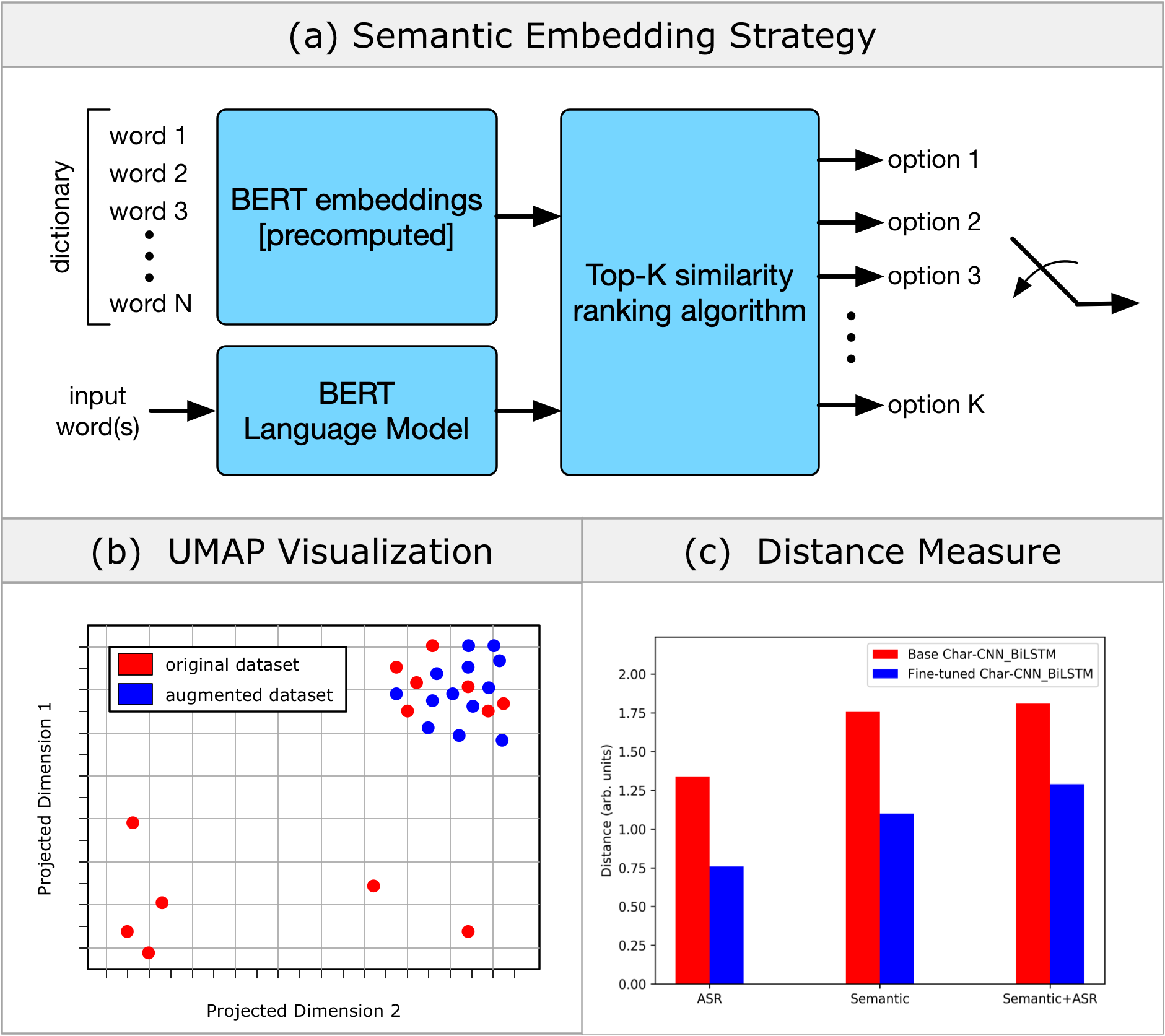}
            \par
        \end{centering}
        \caption{(a) An example semantic augmentation pipeline that leverages a deep neural language model to produce a distance-based word similarity look-up table. (b) Embeddings from the fine tuned Char-CNN-BiLSTM models from both original ASR corrupted text inputs. Projections were performed using the popular nonlinear manifold projection technique UMAP \cite{McInnes-UMA}. (c) The mean euclidean distance between projected PTB data with and without ASR, semantic, and semantic+ASR errors.}
        \label{fig:embed}
    \end{figure}
        For the speech-to-text engine depicted in Fig.~\ref{fig:asr}(a), we utilized the DeepSpeech implementation and model that is open source and available from Mozilla Corporation\footnote{Available at {\tt https://github.com/mozilla/DeepSpeech}}. The model architecture uses several stacked Recurrent Neural Network (RNN) decoders and the Connectionist Temporal Classification (CTC) loss function to handle alignment. 
    
    \subsection{Language model training}
    
        The Char-CNN-BiLSTM language model was trained on the PTB corpus, which consists of $\sim$50K sentences (90-5-5 split), compiled from various telephone speech, news-wire, microphone speech, and transcribed speech data sources. The model was trained using mini-batch ($n=512$) stochastic gradient descent with Nesterov momentum, an initial learning rate of 0.25, and manual learning rate annealing. Typically, language models are trained on batches of highly correlated inputs, in which each input sentence is unraveled to produce $T$ training examples, where $T$ is the number of words in the input sequence. While computationally convenient, this sampling strategy produces samples that are not independently drawn. Our training procedure eliminates this dependence by randomly selecting only one word from each text sample per epoch; this is less efficient computationally, but produces smoother gradients.
        
        Table 1 shows the perplexity values of the Char-CNN-BiLSTM language model on data with and without ASR and semantic perturbations. The baseline Char-CNN-BiLSTM was trained on uncorrupted PTB data, and has a test perplexity of 37.49. The perplexity of this model on ASR, semantic, and semantic+ASR corrupted data, however, is significantly higher (+55.80, +110.66, +133.58, respectively), which demonstrates its inability to generalize to text containing phonetic and semantic noise. Through fine tuning using the Telephonetic framework, we observe a sharp drop in perplexity on the ASR, semantic, and semantic+ASR corrupted test sets relative to the baseline model (-43.76, -87.55, and -101.25, respectively), and only a marginal increase on the original test set (+5.02, +8.06, and +6.93, respectively).
        
        While semantic perturbations appear to have a larger effect on language model perplexity than those from ASR, the effects of ASR perturbations are clear from Fig.~\ref{fig:embed}(b), which shows a 2D projection of the embeddings produced from the baseline and fine-tuned models on original and ASR corrupted text inputs generated from the string {\tt the quick brown fox jumped over the lazy dog}. The ASR augmentation clearly produces decoded text that diverges from the distribution of natural language found in the PTB dataset, as measured in the Char-CNN-BiLSTM latent space. Importantly, we observe that fine tuning on ASR corrupted data provides more robustness to those inputs than does the baseline model trained on only the original data. Figure ~\ref{fig:embed}(c) quantifies the variation in these projections before and after fine tuning, showing the euclidean distance between fine tuned model projections produced from baseline and corrupted inputs is lower than those prodcued by the baseline Char-CNN-BiLSTM model. The projections produced by the fine tuned Char-CNN-BiLSTM models are less sensitive to these noise sources than the baseline.
        
    \begin{table}[th]
      \caption{The performance of the Char-CNN-BiLSTM model on English Penn Treebank data with and without ASR and semantic noise. For reference, the current state-of-the-art perplexity on the PTB test set is OpenAI's GPT-2 \cite{Radford19-LMA} (35.76). The previous state-of-the-art for a non-transfer learned language model was 46.54 \cite{Gong2018FRAGEFW}.}
      \label{tab:example}
      \centering
      \begin{tabular}{ccccc}
        \toprule
         & {\bf Fine tuned on} & {\bf Tested on} & {\bf PPL (valid / test)} \\
        \midrule
        & None              & Baseline          & \textbf{40.20 / 37.49} \\
        & None              & ASR               & 110.02 /  92.85 \\
        & None              & Semantic          & 178.55 / 147.71 \\
        & None              & ASR + Semantic    & 206.00 / 170.73 \\
        & ASR               & Baseline          &  46.14 /  42.07 \\
        & ASR               & ASR               &  52.86 /  49.09 \\
        & Semantic          & Baseline          &  49.61 /  45.11 \\
        & Semantic          & Semantic          &  62.87 /  60.16 \\
        & ASR + Semantic    & Baseline          &  47.97 /  43.98 \\
        & ASR + Semantic    & ASR + Semantic    &  73.56 /  69.52 \\
        \bottomrule
      \end{tabular}
    \end{table}

\section{Conclusions}
    The telephonetic augmentation framework, which leverages a number of recent advances in deep learning, enables state-of-the-art language models trained on only written text datasets to transfer gracefully to handling speech domain inputs corrupted by ASR errors. Empirically, we found that fine tuning on a dataset comprising a combination of semantic and phonetic data perturbations enables character level language models to generalize to out-of-sample inputs containing ASR structured noise. It is also worth noting that these data augmentation strategies come free of labeling cost; BERT, DeepSpeech, WaveGlow, and Tacotron 2 can be leveraged either out-of-the-box or by training/fine tuning on publicly available datasets with no manual labeling required. In addition to the Telephonetic data augmentation strategy presented in this manuscript, we observe a sizeable performance improvement on the language modeling task (PPL reduced by $\sim$20$\%$ with respect to \cite{Gong2018FRAGEFW}) when using character-level model in combination with probabilistic masking and independent mini-batch sampling. This training strategy is inspired by the process that was used to train BERT, and to our knowledge, the baseline perplexity of 37.49 is the lowest perplexity achieved by a non transfer-learned model to date. 

\section{Acknowledgements}

    The authors would like to thank Corey Fyock, Jeremy Doll, Kavita Hundal, Margaret Mayer, Mathias Menasi, Oluwatobi Olabiyi, and Varun Singh for their helpful insight.
    


\end{document}